\documentclass[prb,twocolumn,aps,showpacs,fixfloats]{revtex4}
\usepackage{graphicx}
\usepackage{bm}
\usepackage{amsmath,amssymb}
\usepackage{subfigure}
\usepackage{float}
\usepackage{latexsym}
\usepackage{color}
\usepackage{enumerate}
\usepackage{pdfpages}
\usepackage{tikz}
\usepackage{hyperref}
\usepackage{graphicx}
\usepackage{bm}
\usepackage{amssymb} 
\usepackage{amsmath}
\usepackage{subfigure}
\newcommand{\s}{\scriptscriptstyle}
\usepackage{relsize}
\begin{document}

\title{Long-living excited states of a 2D diamagnetic exciton}

\author{R. E. Putnam, Jr. and M. E. Raikh}

\affiliation{ Department of Physics and Astronomy, University of Utah, Salt Lake City, UT 84112}

\begin{abstract}
Hydrogenic excited states of a 2D exciton are degenerate. In the
presence of a weak magnetic field, the $S$-states with a zero momentum of
the center of mass get coupled to the $P$-states with finite momentum of the
center of mass. This field-induced
coupling leads to a strong modification of the dispersion branches
of the exciton spectrum. Namely, the lower branch acquires a shape
of a ``mexican hat" with a minimum at a finite momentum. At certain magnetic field,
exciton branches exhibit a linear crossing, similarly to the
spectrum of a 2D electron in the presence of spin-orbit coupling. While
spin is not involved, degenerate $S$ and $P$ states play the role of the
spin projections. Lifting of degeneracy due to diamagnetic shifts and
deviation of electron-hole attraction from purely Coulomb suppresses the
linear crossing.

\end{abstract}

\maketitle

\section{Introduction}
Diamagnetic exciton, a bound state of electron and hole in a magnetic field, was studied theoretically and experimentally\cite{Seisyan} for many years.
Originally, the attention was focused on the bulk semiconductors.\cite{Gorkov}
Later, the
interest has shifted to the two-dimensional systems.\cite{Lozovik,Kallin,Macdonald1986,Butov,Zimmermann}
It was established that, for
interband excitons\cite{Lozovik} as well as for inter-Landau-level excitons,\cite{Kallin} that the exciton dispersion law has a local minimum at momenta of the order of the inverse
magnetic length.

\begin{figure}
\includegraphics[width=110mm]{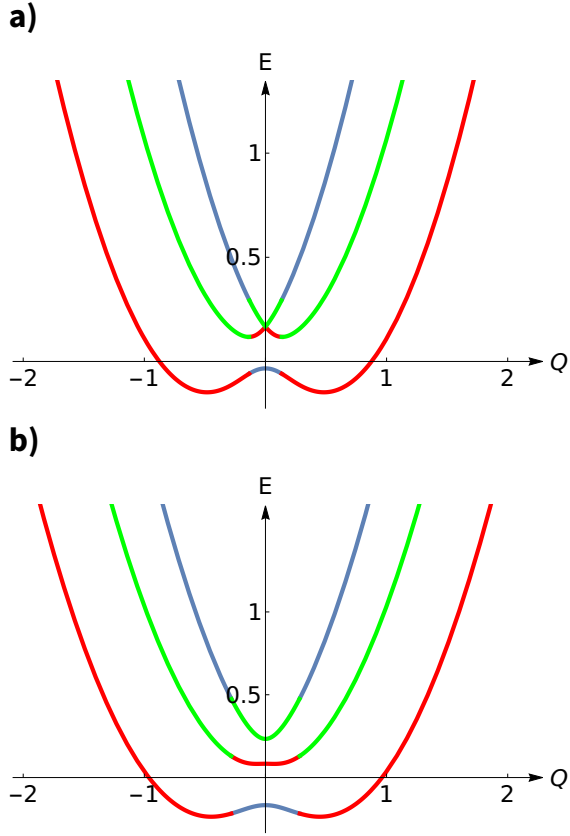}
\caption{(Color online) Three branches of $n=2$ exciton spectrum
are plotted using Eq. (\ref{ThreeBranches}). Different colors
correspond to different solutions of the cubic equation  Eq. (\ref{closed}).
Energy is measured in the units of
$\varepsilon_0$ defined by Eq. (\ref{varepsilon}), while
the momentum $Q$ is measured in the units of
$\left(\frac{2M\varepsilon_0}{\hbar^2}\right)^{1/2}$.
Plot a) corresponds to the resonance $E_{SP} = \hbar\Omega_{-} = \frac{\varepsilon_0}{10}$ for which the dispersion is linear as predicted by Eq. (\ref{crossing}).
  Plot b) corresponds to $E_{SP} = \frac{\varepsilon_0}{20}$ and $\hbar\Omega_{-}= \frac{\varepsilon_0}{5}$. 
  The structure of the spectrum corresponds to the limiting case Eq. (\ref{branches1})
   with a well pronounced minimum in the lower branch.}
\label{F1}
\end{figure}

Recently,
\cite{Zeeman,CrookerNano,Veligzhanin,Crooker,Luminescence,Luminescence1,Potemski1,Potemski2,report} the exciton spectroscopy in a perpendicular magnetic field was applied to the novel van der Waals monolayers.
These materials host a series of exciton Rydberg states corresponding to
the principal quantum number $n=1, 2, ...$. This property is a consequence of strong Coulomb interaction resulting from the reduced dielectric screening.
Experimental spectra  in Refs.
\onlinecite{Zeeman,CrookerNano,Veligzhanin,Crooker,Luminescence,Luminescence1,Potemski1,Potemski2,report}
were interpreted as diamagnetic shifts of the $S$-states
of the exciton. The growth of $\Big[E_n(B)-E_n(0)\Big]$ with magnetic field, ${\bf B}$, gets faster
with increasing $n$. Note, that the
$P$-states of the exciton do not show up in the absorption experiments since
the matrix element between the vacuum and the $P$-states is zero. However,
$P$-states can manifest themselves in luminescence.

In the present paper we consider excited states of a diamagnetic exciton in a {\em weak} magnetic field.
Our main finding is that a non-quantizing magnetic field still affects strongly the
exciton dispersion. The underlying reason is that, due to accidental degeneracy of
the excited states, the motion of the center of mass in a finite field
couples different states of the internal motion.
It is this coupling, together with lifting of the accidental degeneracy due to
deviation from Coulomb attraction at short distances,\cite{Rytova,Keldysh}
that modifies the dispersion law even in a weak  magnetic field.  We demonstrate that such a modification
can give rise to the {\em loop of extrema} in the dispersion law.
In turn, this loop of extrema leads to anomalous broadening of
the exciton absorption line. Another consequence is that, with a minimum in the dispersion law, an
exciton is trapped by arbitrarily weak impurity. Since recombination of a trapped
exciton with rapidly precessing center of mass
requires a big momentum transfer, this state is {\em long-lived}.

\begin{figure}
\includegraphics[width=100mm]{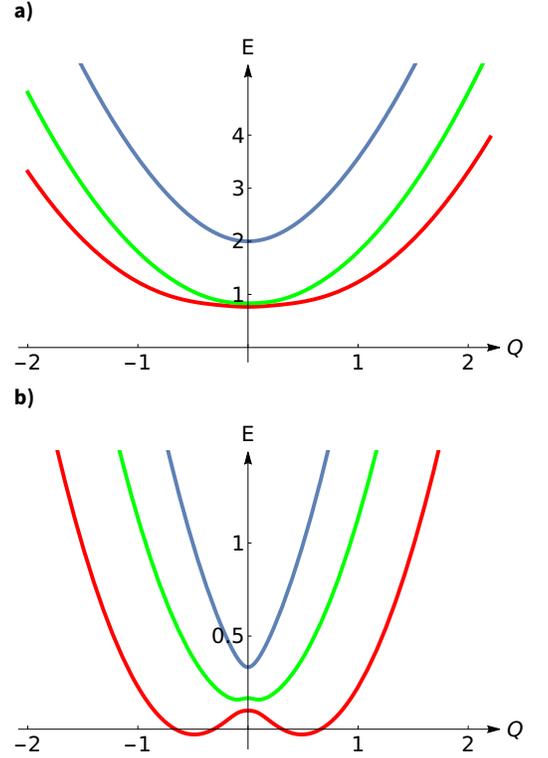}
\caption{(Color online)
Three branches of $n=2$ exciton spectrum
are plotted using Eq. (\ref{ThreeBranches}).
Energy and momentum are measured in the same units as
in Fig. 1. Both panels correspond to $\hbar\Omega_{-} =\frac{\varepsilon_0}{30} $, when two
of the branches are almost degenerate at $Q=0$. According to Eq. (\ref{quadratic}), all three
effective masses are positive for $E_{SP}>\varepsilon_0$, while for $E_{SP}<\varepsilon_0$ the effective
mass of the lower branch is negative. This is illustrated by plots $E_{SP} = \frac{6\varepsilon_0}{5}$, Panel (a)
and $E_{SP} = \frac{\varepsilon_0}{10}$, Panel (b).}
\label{F2}
\end{figure}

\section{Dispersion law of a 2D diamagnetic exciton with $n=2$}

We start with a standard Hamiltonian of diamagnetic exciton\cite{Gorkov,Lozovik,Butov,Zimmermann}
\begin{align}
\label{Hamiltonian}
&\hat{H} = \frac{1}{2m_e}\Big[-i\hbar \nabla_e + \frac{e}{c}{\bf A}({\bm r}_e)\Big]^2 \nonumber\\
&+\frac{1}{2m_h}\Big[-i\hbar\nabla_h - \frac{e}{c}{\bf A}({\bm r}_h)\Big]^2 - \frac{e^2}{\kappa\big\vert {\bf r_e} - {\bf r_h}\big\vert}.
\end{align}
Here $m_e$ and $m_h$ are the masses of electron and hole, $\kappa$ is the dielectric constant. Vector
potential ${\bm A}({\bm r})$ is defined as ${\bm A}({\bm r})=\frac{1}{2}{\bm B}\times {\bm r}$.

Upon introducing the center of mass and the relative coordinates
\begin{equation}
\label{variables}
{\bf R} = \frac{m_e{\bf r_e} + m_h{\bf r_h}}{m_e + m_h},~~{\bf r}={\bf r}_e-{\bf r}_h,
\end{equation}
the Hamiltonian Eq. (\ref{Hamiltonian}) acquires the form
\begin{equation}
\label{Hamiltonian'}
\hat{H}=\hat{H}_{cm}({\bm R})+\hat{H}_{rel}({\bm r})+\hat{H}_c({\bm r},{\bm R}),
\end{equation}
where $\hat{H}_{cm}({\bm R})$ describes the motion of the center of mass,
$\hat{H}_{rel}({\bm r})$ describes the relative motion, and  $\hat{H}_c({\bm r},{\bm R})$
describes their coupling. Analytical expressions for the three Hamiltonians are the following
\begin{align}
&\hat{H}_{cm}=-\frac{\hbar^2}{2M}\Delta_{\bf R},\\
&\hat{H}_{rel}=-\frac{\hbar^2}{2\mu}\Delta_{\bf r}-\frac{e^2}{\kappa r}+\frac{\left(eBr\right)^2}{8\mu c^2}\nonumber\\
&+
i\hbar\Bigl(\frac{eB}{2c}\Bigr)\Bigl(\frac{1}{m_e}-\frac{1}{m_h}\Bigr)\left(y\partial_x - x\partial_y\right),\\
&\hat{H}_c=i\hbar\frac{eB}{Mc}\left(y\partial_X - x\partial_Y\right),
\end{align}
where $M=m_e+m_h$ is the net mass and $\mu=m_em_h/(m_e+m_h)$ is the reduced mass
of the exciton.

In the absence of magnetic field, the states $n=2$ of the exciton are triply degenerate.
At finite field, the degeneracy is lifted by the third and the fourth terms in $\hat{H}_{rel}$.
Due to the third term, diamagnetic shifts are different for the $S$ and $P$-states.
Fourth term results in the repulsion of the two $P$-states. In addition, the $S$ and
$P$-states are coupled to each other via the center-of-mass motion. This coupling
is promulgated by the Hamiltonian $\hat{H}_c$.
Our main point is that the interplay of the three
effects leads to a nontrivial modification of the dispersion law of the exciton.
In contrast to Refs. \onlinecite{Lozovik,Kallin}, this  nontrivial modification takes
place at low fields, so that all three effects can be taken into account {\em perturbatively}.

Explicit form of the $n=2$ wavefunctions is the following
\begin{align}
\label{2S}
&\psi_0({\bm r}) = C_0\Big(1-\frac{r}{l}\Big)\exp\left[-\frac{r}{2l}\right],\\
&\psi_x({\bm r}) = C_x \left(\frac{x}{l}\right) \exp\left[-\frac{r}{2l}\right],~
\psi_y({\bm r}) = C_y \left(\frac{y}{l}\right) \exp\left[-\frac{r}{2l}\right],
\end{align}
where $l$ is expressed via the Bohr radius $a_B=\frac{\hbar^2\kappa}{\mu e^2}$
as $l=\frac{3}{4}a_B$. The wavefunctions Eq. (\ref{2S}) correspond to the binding
energy $\frac{4}{9}E_B$, where $E_B=\frac{\mu e^4}{2\hbar^2\kappa^2}$ is the Bohr energy.
Normalization constants are the same for all three functions
\begin{equation}
   C_0= C_x = C_y =\frac{1}{\left(6\pi\right)^{1/2}l}.
\end{equation}
With the help of the wavefunctions Eq. (\ref{2S}) we calculate the diamagnetic shifts
\begin{align}
\label{shifts}
    E_S =  \frac{e^2B^2}{8\mu c^2}\Big<r^2\Big>_{S},~~E_P = \frac{e^2B^2}{8\mu c^2}\Big<r^2\Big>_{P}.
\end{align}
Elementary integration yields
\begin{equation}
\label{shifts1}
\Big<r^2\Big>_{S}=26l^2,~~~\Big<r^2\Big>_{P}=20l^2.
\end{equation}
Thus, the shifts of $2S$ and $2P$ states are related as $E_S = 1.3E_P$,
with $2S$ state being higher.

Next we evaluate the coupling coefficient between the
states $\psi_0$ and $\psi_x, \psi_y$. As follows from the
form of the Hamiltonian $\hat{H}_c$, this matrix element
contains

\begin{equation}
\label{xsp}
x_{SP}=\int d{\bf r}\psi_0({\bf r})x \psi_x({\bf r}).
\end{equation}
Using the wavefunctions  Eq. (\ref{2S}), we find
\begin{equation}
\label{xsp1}
x_{SP}=C_0^2\int\limits_0^{\infty}dr r\int_0^{2\pi} d\phi~ \cos^2\phi~ \frac{r^2}{l}\left(1-\frac{r}{l}\right)\exp\left[-\frac{r}{l}   \right].
\end{equation}
Angular integration yields $\pi$. Performing the radial
integration, we find
\begin{equation}
\label{xsp2}
x_{SP}=\pi C_0^2\int\limits_0^{\infty}dr~\frac{r^3}{l}\left(1-\frac{r}{l}\right)\exp\left[-\frac{r}{l}   \right]=-3l.
\end{equation}

Since the Hamiltonian $\hat{H}_c$ contains $\partial_X$ and $\partial_Y$,
the coupling between the $S$ and $P$-states depends on the motion of the
center of mass. If the center of mass moves with momentum, $Q$, the
general form of the exciton wavefunction can be  presented as
a linear combination

\begin{align}
\label{Psi-2}
    \Psi({\bf r},{\bf R}) = \exp[i{\bf QR}]\Big[A_0\psi_0({\bm r}) + A_x\psi_x({\bm r})
    +A_y\psi_y({\bm r})\Big].
\end{align}
Substituting the form Eq. (\ref{Psi-2}) into the Schr{\"o}dinger equation, ${\hat H}\Psi=E\Psi$, yields the system of equations for the coefficients
$A_0, A_x$, and $A_y$

\begin{align}
\label{S''}
&\Bigg(\frac{\hbar^2Q^2}{2M} +E_S - E\Bigg)A_0- \hbar\Omega_+x_{SP}(Q_yA_x - Q_xA_y) = 0,\\
\label{2}
&\Bigg(\frac{\hbar^2Q^2}{2M} +E_P - E\Bigg)A_x - i\hbar\Omega_{-}A_y
-\hbar\Omega_+Q_y x_{SP}A_0=0,\\
\label{3}
&\Bigg(\frac{\hbar^2Q^2}{2M} +E_P - E\Bigg)A_y + i\hbar\Omega_{-}A_x
-\hbar\Omega_+Q_x x_{SP}A_0=0.
\end{align}
Here we have introduced two magnetic-field-induced  energy scales
\begin{equation}
\label{scales}
\hbar\Omega_{-}=\hbar\frac{eB}{2c}\Big(\frac{1}{m_e}-\frac{1}{m_h}\Big),
~\hbar\Omega_{+}=\hbar\frac{eB}{Mc}.
\end{equation}

Expressing $A_x$, $A_y$ from Eqs. (\ref{2}), (\ref{3})
and substituting them into Eq. (\ref{S''}), we arrive at the closed
equation for the dispersion $E(Q)$

\begin{flalign}
\label{closed}
    &\Bigg(\frac{\hbar^2Q^2}{2M} + E_S - E\Bigg)\Bigg[\Bigg(\frac{\hbar^2Q^2}{2M} + E_P - E\Bigg)^2 -(\hbar\Omega_-)^2\Bigg] \nonumber\\
    & = \Big(\hbar\Omega_+x_{SP}\Big)^2Q^2\Bigg(\frac{\hbar^2Q^2}{2M} + E_P - E\Bigg).
\end{flalign}
Three solutions of Eq. (\ref{closed}) determine three branches of the
exciton dispersion. We analyze this dispersion in the next Section.
\section{Limiting Cases}

\subsection{$m_e=m_h$}
It is seen from Eq. (\ref{scales}) that the scales $\Omega_{-}$ and $\Omega_{+}$ are strongly different under the condition $|m_e-m_h|\ll \mu$.
When this condition is satisfied, we can neglect $\Omega_{-}$ in Eq.~ (\ref{closed}). After that, three branches of the exciton spectrum can be
easily found. While one branch is purely parabolic: $E=E_P+\frac{\hbar Q^2}{2M}$, two other branches satisfy the quadratic equation
\begin{flalign}
\label{closed1}
    \Bigg(\frac{\hbar^2Q^2}{2M} + E_S - E\Bigg)\Bigg(\frac{\hbar^2Q^2}{2M} + E_P - E\Bigg)=\varepsilon_0\frac{\hbar^2Q^2}{2M},
\end{flalign}
where we have introduced the energy
\begin{equation}
\label{varepsilon}
\varepsilon_0=2M\Omega_{+}^2x_{SP}^2.
\end{equation}
Consider the expression for the lower branch
\begin{align}
\label{quadratic}
    E - \frac{\hbar^2Q^2}{2M} = \frac{E_S + E_P}{2}-
    \Bigg[\frac{(E_S - E_P)^2}{4} + \frac{\hbar^2Q^2}{2M}\varepsilon_0\Bigg]^{1/2}.
\end{align}
Our main point is that the spectrum, $E(Q)$, has a minimum for
large enough $\varepsilon_0$.
From Eq. (\ref{quadratic}) we find the position of minimum
\begin{equation}
\label{minimum}
Q_{\text{min}}=
\Big(\frac{M}{2\varepsilon_0\hbar^2}\Big)^{1/2}
\Big[\varepsilon_0^2-\left(E_S-E_P   \right)^2 \Big]^{1/2}.
\end{equation}
Substituting $Q_{\text{min}}$ back into Eq. (\ref{quadratic}),
we find the depth of the minimum
\begin{equation}
\label{minimum}
E(Q_{\text{min}})-E(0)=-\frac{1}{4\varepsilon_0}\Big[E_S-E_P-\varepsilon_0     \Big]^2.
\end{equation}
We see that the minimum emerges when $\varepsilon_0 > \left(E_S-E_P\right)$.
Note that both $\varepsilon_0$ and $\left(E_S-E_P\right)$ depend on magnetic
field as $B^2$. Then the condition of minimum takes the form
\begin{equation}
\label{condition}
x_{SP}^2>\frac{1}{16}\left(\frac{M}{\mu}\right)\Bigg[\Big<r^2\Big>_S   -\Big<r^2\Big>_P\Bigg].
\end{equation}
With the help of Eqs. (\ref{shifts1}), (\ref{xsp2})
the above condition reduces to $M < 24\mu$.
We thus conclude that the minimum in the exciton spectrum
is quite generic.

\subsection{$\hbar\Omega_{-}\gg\left(E_S-E_P\right)$}

In this limit we can neglect the difference, $\left(E_S-E_P\right)$,
in Eq. (\ref{closed}). Then the first branch is still purely parabolic, $E=E_P+\frac{\hbar^2Q^2}{2M}$, while two other branches are given by
\begin{equation}
\label{branches1}
 E - \frac{\hbar^2Q^2}{2M} = E_P\pm
    \Big[\left(\hbar\Omega_{-}\right)^2 + \frac{\hbar^2Q^2}{2M}\varepsilon_0\Big]^{1/2}.
\end{equation}
Similarly to Eq. (\ref{quadratic}) the lower branch develops a minimum
when the condition $\varepsilon_0>2\hbar\Omega_{-}$ is met. Note, that
$\varepsilon_0$ grows with $B$ quadratically, while $\Omega_{-}$ grows
with $B$ linearly. Thus, the minimum can be enforced upon increasing
magnetic field.

\begin{figure}
\includegraphics[width=103mm]{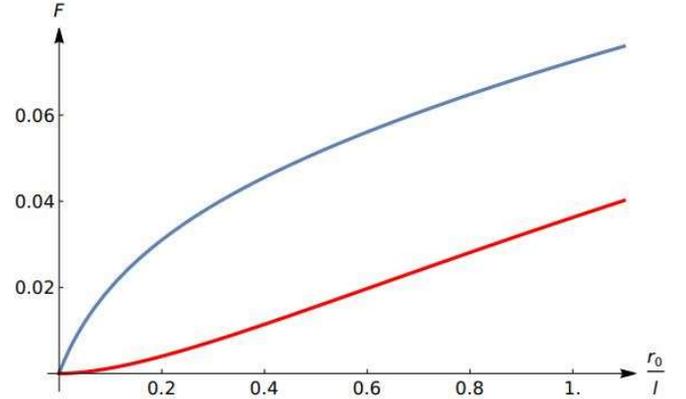}
\caption{(Color online)
 The shifts, $\delta_S$ (blue curve) and $\delta_P$ (red curve),
 of the exciton levels are plotted versus the dimensionless screening parameter, $r_0/l$,
 from Eqs. (\ref{dimensionless1})
and (\ref{dimensionless2}), respectively. At small $r_0/l$ the shift
$\delta_S$ grows linearly with $r_0$, while $\delta_P$ grows quadratically.}
\label{figure1}
\end{figure}

\section{Modification of $E_S$ and $E_P$ by the Keldysh potential}

At short distances between electron and hole their
attraction deviates from purely Coulomb as a result of potential
created by polarization charges created in the 2D plane, see e.g.
Refs.  \onlinecite{Rytova,Keldysh,Cudazzo,Lewenkopf,Dery}.
The modified interaction can be parametrized by a single
parameter, $r_0$, proportional to polarizability, and
has a form

\begin{align}
\label{modified}
    V_{\s eff}(\pmb{r}) = \frac{\pi e^2}{2r_0}\Bigg[H_0\Big(\frac{{r}}{r_0}\Big) - Y_0\Big(\frac{r}{r_0}\Big)\Bigg],
\end{align}
where $H_0$ and $Y_0$ are the Struve and second-kind Bessel functions, respectively. With modified interaction, the accidental degeneracy of
the $n=2$ exciton level is lifted even in the absence of magnetic field.
Since the shifts of the $S$ and $P$-levels are relatively small,
the deviation of $V_{\s eff}(\pmb{r})$ from the Coulomb attraction
can be taken into account perturbatively. Corrections to the energy of $S$ and $P$-states read
\begin{align}
\label{correction}
&\delta_S=\int d{\bf r}\Bigl(V_{\s eff}(r)-\frac{e^2}{\kappa r}\Bigr)\psi_0({\bm r})^2, \nonumber\\
&\delta_P=\int d{\bf r} \Bigl(V_{\s eff}(r)-\frac{e^2}{\kappa r}\Bigr)   \psi_x({\bm r})^2.
\end{align}
It is convenient to evaluate $\delta_S$ and $\delta_P$ in the momentum space.
For this purpose, we present the Fourier transform of $V_{\s eff}(\pmb{r})$ in the form
\begin{equation}
\label{Fourier}
\Phi(q) = -\frac{2\pi e^2}{\kappa q}\Big(1 - \frac{qr_0}{1+qr_0}\Big)
\end{equation}
and treat the second term as a perturbation. Then the expressions for $\delta_S$, $\delta_P$ take the form
\begin{align}
\label{correction1}
\delta_S=\frac{2\pi e^2r_0}{\kappa}\int\frac{d{\bf q}~ q}{1+qr_0}U_0(q),~\delta_P=\frac{2\pi e^2r_0}{\kappa}\int\frac{d{\bf q}~ q}{1+qr_0}U_x({\bf q}),
\end{align}
where $U_0(q)$ and $U_x({\bf q})$ stand for the Fourier transforms
\begin{align}
\label{correction2}
&U_0(q)=\int d{\bf r}\psi_0({\bm r})^2\exp\left(i{\bf q r}   \right),
\nonumber\\
&U_x({\bf q})=\int d{\bf r}\psi_x({\bm r})^2\exp\left(i{\bf q r}   \right).
\end{align}
In the first transform, angular integration of $\exp\left(i{\bf q r}   \right)$ yields the zero-order
Bessel function, $J_0(qr)$. Then the radial integration is performed with the
help of the identity
\begin{align}
\label{Bessel}
    \int\limits_0^{\infty} d{t} J_0(t)\exp\left({-\gamma t}\right) = \frac{1}{(\gamma^2 + 1)^{1/2}}.
\end{align}
The result reads
\begin{equation}
\label{transform1}
U_0(q)=\frac{1-3q^2l^2+q^4l^4}{\left(1+q^2l^2   \right)^{7/2}}.
\end{equation}
Substituting this result into Eq. (\ref{correction2}), we rewrite $\delta_S$ in the dimensionless form

\begin{align}
\label{dimensionless1}
\delta_S=\frac{e^2}{\kappa l}F_S\left(\frac{r_0}{l}\right),~
F_S=\frac{r_0}{l}\int\limits_0^{\infty}\frac{dQ Q}
{1+\frac{r_0}{l}Q}\frac{1-3Q^2+Q^4}{\left(1+Q^2\right)^{7/2}}.
\end{align}
We see that the integral is a function of the dimensionless ratio $r_0/l$.

Angular integration in $U_x({\bf q})$ is less trivial.
This is because both the exponent
$\exp\left(i{\bf q r}\right)$ and  the pre-exponential factor contain the
angle $\phi_{\bf r}$. One has

\begin{flalign}
&U_x({\bf q})
= C^2_0\!\! \int\limits_0^{2\pi}\int\limits_0^{\infty} d{\phi_{\bf r}} dr r  \Big(\frac{r \cos\phi_{\pmb{r}}}{l}\Big)^2\!
\exp\Big(- \frac{r}{l} \Big) \nonumber\\
&\times\exp\Big(iqr \cos(\phi_{{\bf q}} - \phi_{\pmb{r}})\Big).
\end{flalign}
The result of integration contains a constant part and two other parts
proportional to $\cos(2\phi_{\bf q})$ and $\sin(2\phi_{\bf q})$. Only
a constant part contributes to $\delta_P$. To calculate this constant part
it is sufficient to replace $\cos\phi_{\pmb{r}}^2$ by $1/2$.
Subsequent steps are similar to those in calculation of $\delta_S$
and leads to the result

\begin{equation}
\label{dimensionless2}
\delta_P=\frac{e^2}{\kappa l}F_P\left(\frac{r_0}{l}\right),~ F_P=\frac{r_0}{l}\int\limits_0^{\infty}\frac{dQ Q}
{1+\frac{r_0}{l}Q}\frac{1-\frac{3}{2}Q^2}{\left(1+Q^2\right)^{7/2}}.
\end{equation}
\vspace{30mm}
At small $r_0\ll l$ the integral is proportional to $r_0$. This is a
consequence of the fact that the wavefunction of the $P$-state turns
to zero at the origin. The functions $F_S\left(\frac{r_0}{l}\right)$
and $F_P\left(\frac{r_0}{l}\right)$ are plotted in Fig. \ref{figure1}.
Realistic value of $\frac{r_0}{l}$ for transition-metal dichalcogenides
can be estimated e.g. using the data of Ref. \onlinecite{Crooker}.
From the spectroscopic measurements the value $r_0\approx 3.5$nm was
inferred, while the radius of the ground-state wavefunction $a_B\approx 1.5$nm.
Thus, for the first excited state, the ratio $\frac{r_0}{l}$ is $\sim 1$.
\section{General case}
Corrections $\delta_S$, $\delta_P$ are independent of magnetic field, while
$E_S$ and $E_P$ are proportional to $B^2$. Thus, the difference
\begin{equation}
\label{Esp}
E_{SP}=\Big(E_S+\delta_S\Big) -\Big(E_P+\delta_P\Big)
\end{equation}
is a linear function of magnetic field. Two other parameters, $\hbar\Omega_{-}$ and $\varepsilon_0$, which enter into the equation
Eq. (\ref{closed}), are proportional to $B$ and to $B^2$, respectively.
Overall, the evolution of the exciton branches with magnetic field is
quite nontrivial. To examine this evolution, we turn to the analytical
solutions of the cubic equation. The shortest way to arrive to these
solutions is performing the following substitution in Eq. (\ref{closed})
\begin{equation}
\label{substitution}
E=\frac{\hbar^2Q^2}{2M}+\frac{E_{SP}}{3}-
\Bigg[\frac{E_{SP}^2}{3} +\left(\hbar \Omega_{-}\right)^2 +
\varepsilon_0\frac{\hbar^2Q^2}{2M}\Bigg]^{1/2}\eta,
\end{equation}
Then the cubic equation for $\eta$ assumes the form
\begin{equation}
\label{cubic}
\eta^3-\eta +f=0.
\end{equation}
Here the dimensionless parameter $f$
is the following combination of $E_{SP}$, $\hbar\Omega_{-}$,
and $\varepsilon_0$
\begin{equation}
\label{f}
f(Q)=
\frac{E_{SP}\Big[\frac{2}{9}E_{SP}^2-2\left(\hbar \Omega_{-}\right)^2
+\varepsilon_0\frac{\hbar^2Q^2}{2M}\Big]}
{3\Big[\frac{1}{3}E_{SP}^2+\left(\hbar \Omega_{-}\right)^2
+\varepsilon_0\frac{\hbar^2Q^2}{2M}\Big]^{3/2}}.
\end{equation}
Three solutions of Eq. (\ref{cubic}) can be easily expressed
via the phase $\varphi$ defined as
\begin{equation}
\label{varphi}
\varphi(Q)=\arctan \left\{\frac{1}{f(Q)}\left(\frac{4}{27}-f(Q)^2\right)^{1/2} \right\}.
\end{equation}
Analytical form of these solutions is the following
\begin{align}
\label{ThreeBranches}
&\eta_0=-\frac{2}{\sqrt{3}}\cos\Big(\frac{\varphi}{3}\Big),~
\eta_{\pm}=-\frac{2}{\sqrt{3}}\cos\Big(\frac{\varphi}{3}\pm \frac{2\pi}{3}\Big).
\end{align}
We see that the character of solutions changes as $f$ passes through
the value $\frac{2}{3^{3/2}}$, when the phase passes through zero.
In particular, for $Q=0$ the value $f=\frac{2}{3^{3/2}}$ is achieved
under the condition $E_{SP}=\hbar\Omega_{-}$. In fact, this condition
corresponds to the linear crossing of the two branches.
Introducing the deviation
\begin{equation}
\label{deviation}
\hbar\Omega_{-}-E_{SP}=\Delta
\end{equation}
and expanding Eq. (\ref{closed}) at small $Q$ and $\Delta$, we find
behavior of two close branches near the condition $E_{SP}=\hbar\Omega_{-}$
\begin{equation}
\label{crossing}
E=\frac{\hbar^2Q^2}{2M}+\frac{\Delta}{2}\pm \Bigg[\left(\frac{\Delta}{2} \right)^2+\frac{\varepsilon_0}{2}\left(\frac{\hbar^2Q^2}{2M}    \right)     \Bigg]^{1/2}.
\end{equation}
We see that a gap of a width, $\Delta$, opens in the spectrum at finite $\Delta$.

Finally, by expanding  Eq. (\ref{closed}), we find the behavior of all three branches near $Q=0$
\begin{align}
\label{masses}
&E=E_{SP}+\frac{\hbar^2Q^2}{2M}\Bigg[1+\frac{\varepsilon_0E_{SP}}{E_{SP}^2-
\left(\hbar\Omega_{-}   \right)^2}   \Bigg], \\
\label{masses1}
&E=-\hbar\Omega_{-} +\frac{\hbar^2Q^2}{2M}\Bigg[1-\frac{\varepsilon_0}
{2\left(E_{SP}+\hbar\Omega_{-}   \right)}  \Bigg], \\
\label{masses2}
&E=\hbar\Omega_{-}+\frac{\hbar^2Q^2}{2M}\Bigg[1-\frac{\varepsilon_0}
{2\left(E_{SP}-\hbar\Omega_{-}   \right)}  \Bigg].
\end{align}
We conclude that negative effective mass at $Q=0$ emerges at
$\left(E_{SP}\pm \hbar\Omega_{-}\right)<\frac{1}{2}\varepsilon_0$.
Approaching of denominators in Eqs. (\ref{masses1}), (\ref{masses2}) to zero
signals the proximity to the exciton resonance.

\section{Discussion}
Our most nontrivial finding is the exciton resonance originating
from the accidental degeneracy of the hydrogen-like excited levels.
The resonant condition reads $E_{SP}=\pm \hbar\Omega_{-}$. It corresponds to
the magnetic field at which the mismatch of $S$ and $P$ exciton levels
due to diamagnetic shift as well as due to field-independent Keldysh potential,
is equal to the field-induced splitting of the degenerate $P$-states.
All three branches of the bare exciton spectrum ($S$-branch and two $P$-branches)
are involved into the formation of the resonance.
 Under the resonant condition, two branches of the modified
spectrum cross {\em linearly}, as illustrated in Fig. \ref{F1}a.
Away from the resonance, see Figs. \ref{F1}b and \ref{F2}b,
the lowest branch of the spectrum evolves into a ``mexican hat" as
predicted by  Eqs. (\ref{branches1}), (\ref{quadratic})
for the limiting cases $E_{SP}\ll \hbar\Omega_{-}$ and $E_{SP}\gg \hbar\Omega_{-}$,
respectively. The main condition for the strong mixing of the $S$ and $P$-branches
is $\varepsilon_0>\left(E_{SP}+\hbar\Omega_{-}\right)$. We have demonstrated that,
neglecting the Keldysh shift, and for $m_e=m_h$ this condition is satisfied.
Physically, it requires
that the matrix element, $x_{SP}$, defined by Eq.~(\ref{xsp}) is big enough. When
this condition is violated, the effective masses for all three branches
of the modified spectrum are positive, as illustrated in Fig. \ref{F1}a.

With regard to physical consequences of the spectrum modification
into a mexican hat, it is known\cite{Efros,Galstyan,Mkhitaryan,Voloshin}
that the density of states near the minimum behaves as $\big(E-E_{min}\big)^{-1/2}$, i.e.
it has a {\em one-dimensional character}. As a result, even a weak attractive impurity can trap
an exciton. The wave function of the trapped exciton is centered around $Q_{min}$ in the
momentum space. With $Q_{min}$ exceeding the momentum of a photon required for radiative
recombination, these trapped states are long-lived.

Throughout the paper we considered the simplest model of diamagnetic
exciton without account for the valley effects which lead to the polarization
dependence of the optical absorption as well as the spin-orbit effects leading
to the dark-bright exciton mixing.\cite{Durnev} We have also assumed that the coupling of the $S$ and $P$-exciton states is exclusively due to magnetic field and not by the
peculiarities of the bandstructure of dichalcogenide monolayers.\cite{Glazov}

While the optical absorption probes only the $S$-states of the exciton,
finite-momentum states can be probed in the microcavity setting.\cite{microcavity}
Another consequence of the field-induced modification of the exciton spectrum
is anomalous broadening of
$n=2$ absorption line. The underlying mechanism\cite{semimagnetic,WithDisorder}
is the elastic scattering of $Q=0$ state into the states with
finite center-of-mass momentum, $Q$. This scattering is enabled by the
mexican-hat shape of the spectrum.

\section{Acknowledgements}

\vspace{2mm}

The work was supported by the Department of Energy,
Office of Basic Energy Sciences, Grant No.  DE- FG02-
06ER46313.

\end{document}